\documentclass[]{aa} 
\usepackage{natbib}
\bibpunct{(}{)}{;}{a}{}{,} 
\usepackage{graphicx}
\usepackage{txfonts}
\begin{document}
   \title{ Evidence against the young hot-Jupiter around BD\,+20\,1790\thanks{Based on observations 
   collected with CORALIE echelle spectrograph mounted on the Euler 1.2\,m Swiss telescope at La Silla, Chile.
}}

   \author{P. Figueira\inst{1}, 
          M. Marmier\inst{1},
        X. Bonfils\inst{1,2},
        E. di Folco\inst{3},
        S. Udry\inst{1},
        N. C. Santos\inst{4},
        C. Lovis\inst{1},
        D. M\'{e}gevand\inst{1},
        C. H. F. Melo\inst{5},
        F. Pepe\inst{1},
        D. Queloz\inst{1},
        D. S\'{e}gransan\inst{1},        
        A. H. M. J. Triaud\inst{1},
                   \and 
        P. Viana Almeida\inst{4,5}
          }

   \institute{Observatoire Astronomique de l'Universit\'{e} de Gen\`{e}ve, 51 Ch. des Maillettes, 
    - Sauverny - CH1290, Versoix, Suisse\\
     \email{pedro.figueira@unige.ch}
     	\and
	Laboratoire d'Astrophysique, Observatoire de Grenoble, Universit\'{e} J. Fourier, CNRS (UMR5571), BP 53, 38041 Grenoble, Cedex 9, France 
	\and
	 Laboratoire AIM, CEA Saclay-Universit\'{e} Paris Diderot-CNRS, DSM/Irfu/Service d'Astrophysique, 91191 Gif-sur-Yvette, France
     	\and
	Centro de Astrof\'{i}sica, Universidade do Porto, Rua das Estrelas, 4150-762 Porto, Portugal
	\and
	ESO, Alonso de Cordova 3107, Casilla 19001, Vitacura, Santiago, Chile 
	}

   \date{}

  \abstract{The young active star BD\,+20\,1790 is believed to host a substellar companion, revealed by radial-velocity measurements that detected the reflex motion induced on the parent star. }{A complete characterisation of the radial-velocity signal is necessary in order to assess its nature.}{We used CORALIE spectrograph to obtain precise ($\sim$10\,m/s) velocity measurements on this active star, while characterizing the bisector span variations. Particular attention was given to correctly sample both the proposed planetary orbital period, of 7.8\,days, and the stellar rotation period, of 2.4\,days. }{A smaller radial-velocity signal (with peak-to-peak variations $<$500\,m/s) than had been reported previously was detected, with different amplitude on two different campaigns. A periodicity similar to the rotational period is found on the data, as well as a clear correlation between radial-velocities and bisector span. This evidence points towards a stellar origin of the radial-velocity variations of the star instead of a baricentric movement of the star, and repudiates the reported detection of a hot-Jupiter.}{}

\authorrunning{P. Figueira et al.}
\titlerunning{Evidence against the young hot-Jupiter around BD\,+20\,1790}

   \maketitle
%

\section{Introduction}

Since the discovery of the first exoplanet around 51\,Peg by \cite{1995Natur.378..355M}, the radial-velocity method (RV) has established itself as the workhorse for exoplanet detections. It allowed the detection of more than 3/4 of all planets we know and was instrumental in shaping the body of knowledge we gathered on the subject \citep{2007ARA&A..45..397U}. 

Still, many open questions remain. For instance the planetary formation process is a subject of great debate. As a consequence, a positive detection of a planet around a young star would be highly valuable; it would put a stringent upper limit on the time scale of planet formation, providing a strong observational constraint for the modeling.
In order to address this question, several RV surveys started targeting young objects, only to find that extrasolar planets around these hosts were very rare \citep{2007ApJ...660L.145S}. On top of that, since young stars exhibit high photospheric and chromospheric activity, these surveys are plagued by stellar activity effects. The detection of an RV signature rooted on atmospheric phenomena is very common and false planetary detections around very young stars were provided by \cite{2008ApJ...687L.103P} and \cite{2008A&A...489L...9H}, among several others.

Very recently \cite{2009arXiv0912.2773H} provided compelling evidence for a planet around the young active star BD\,+20\,1790. RV observations were associated with a massive hot-Jupiter orbiting the star. Aiming primarily at activity characterisation, several high-frequency RV data sets had been obtained, scattered, spanning six years. The main limitations of the data used were the inadequacy of the time sampling to a planetary search campaign and the low RV precision of the measurements. Still, an extensive activity analysis and careful discussion were presented. The authors showed that some spectral/activity indicators exhibited variation with a time scale of the rotation period (2.8\,days) but none on a time scale of the reported RV variation (7.8\,days). This backed the planetary hypothesis, which was considered as the best explanation for the RV variations.


Following the announcement of the putative planet we started an intensive RV campaign on BD\,+20\,1790 with the CORALIE spectrograph. We analyze our measurements in this paper. In Sect. 2 we present an overview of the parameters both of the star and of the putative planet. In Sect 3 we describe the results of our campaign and discuss these in Sect. 4. We finish by stating our conclusions in Sect. 5.


\section{An overview of the star and planet properties}

The star BD\,+20\,1790 is a K5Ve star with magnitude V=9.9. It has an effective temperature of 4410\,K and an age of 35-80\,Myr. For the provenance of these values and a more detailed description of the host star properties the reader is referred to \cite{2009arXiv0912.2773H}. Of particular importance to us are the measured photometric period of 2.801$\pm$0.001\,days and the $v.\sin{i}$ of 10.03$\pm$0.47\,km/s.

The high activity level of the star is attested by the detection of transient absorption features on the spectra and strong optical flare events. The duration of both phenomena was of several hours. The authors estimated the flare occurrence corresponded to 40\% of the total observing time and showed that the presence of a flare was associated with an increase on the scatter of the bisector span values.

The RV presented by \cite{2009arXiv0912.2773H} showed a peak-to-peak amplitude of $\sim$2\,km/s. The periodogram of their data yielded a 7.78 days peak with a false-alarm probability of 0.35\%. The fit of a keplerian led to a semi-amplitude of 0.8-0.9\,km/s, depending on the allowed range for orbital eccentricity. This value was much higher than the individual uncertainty on the measurements, of 100-200\,m/s.

The authors analyzed the different spectroscopic activity indicators  -- H$_{\alpha}$, H$_{\beta}$, Ca II IRT and Ca II H \& K -- and found that none showed significant variation with a periodicity similar to that of the proposed orbit, and only H$_{\alpha}$ showed a periodicity similar to that of the photometric period. The authors also concluded that the RV variations induced by a star spot would have an amplitude roughly 2 times smaller than the measured one and thus could not explain the measured RV. The bisector signal showed no correlation with RV, and a planetary origin was attributed to the signal.


\section{Radial Velocity measurements \& Analysis}

We obtained a set of 28 CORALIE RV measurements, spanning 55 days. CORALIE is the fiber-fed echelle spectrograph mounted at the Swiss telescope at La Silla observatory. High-precision RV measurements are obtained by cross-correlating the spectra with a template mask \citep{1996A&AS..119..373B, 2002A&A...388..632P}. The observations are reduced online, allowing a real-time calculation of the RV and photon noise estimation. Previous campaigns showed that CORALIE can reach a long-term precision of 5-6\,m/s \citep[e.g.][]{2009arXiv0908.1479S}. We also calculated the bisector velocity span of the cross-correlation function, following the procedure described in \cite{2001A&A...379..279Q}. To do so we calculate the line that bisects the cross-correlation function; the {\it bisector top} and the {\it bisector bottom} are defined as the average bisector values for the ranges of 10 to 40\% and 60 to 90\% of the line depth, respectively. The bisector span (henceforth BIS) is then the inverse of the slope of the line that connects the {\it bisector top} to the {\it bisector bottom}.

\begin{figure}
\includegraphics[width=9cm]{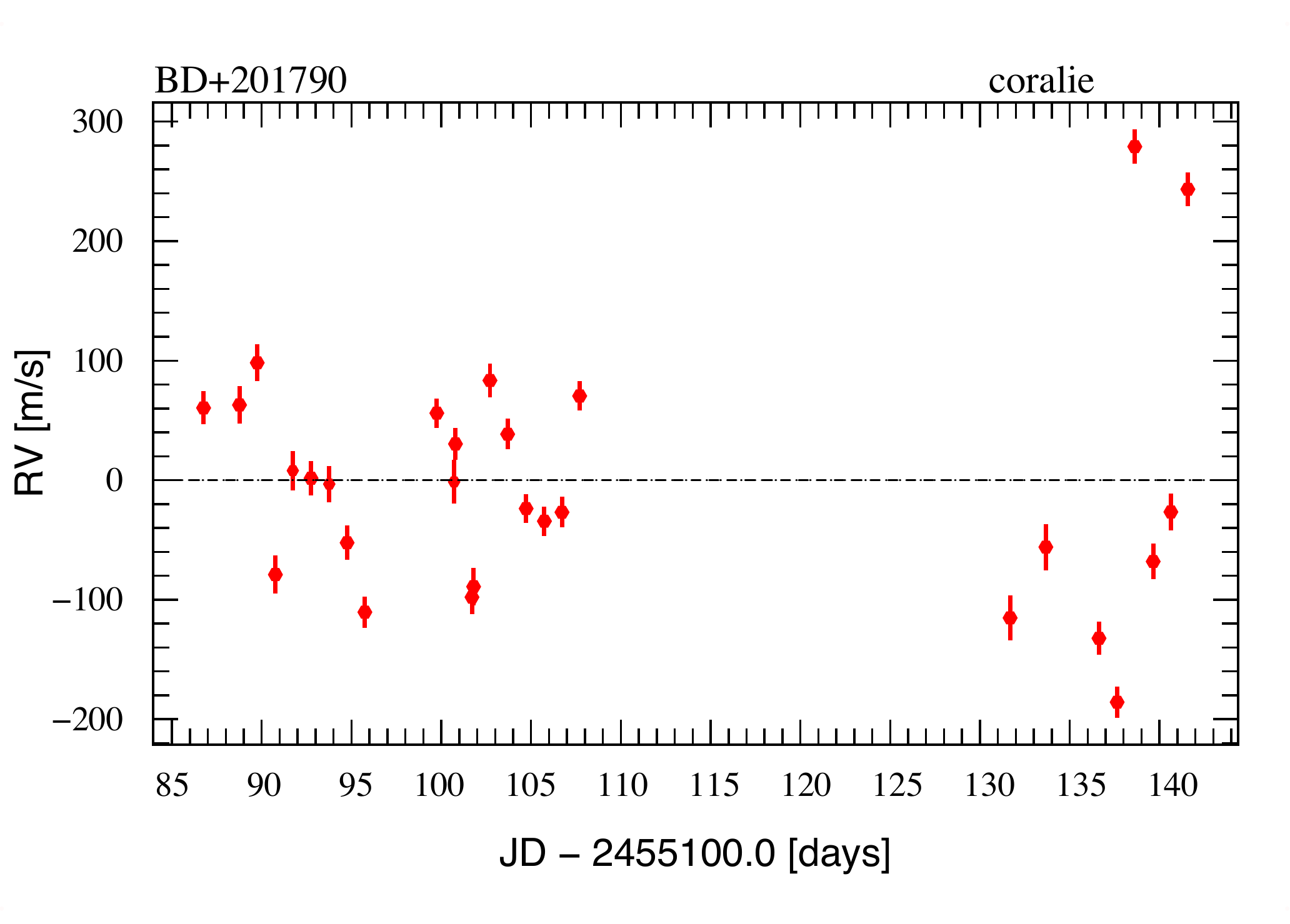}
\includegraphics[width=9cm]{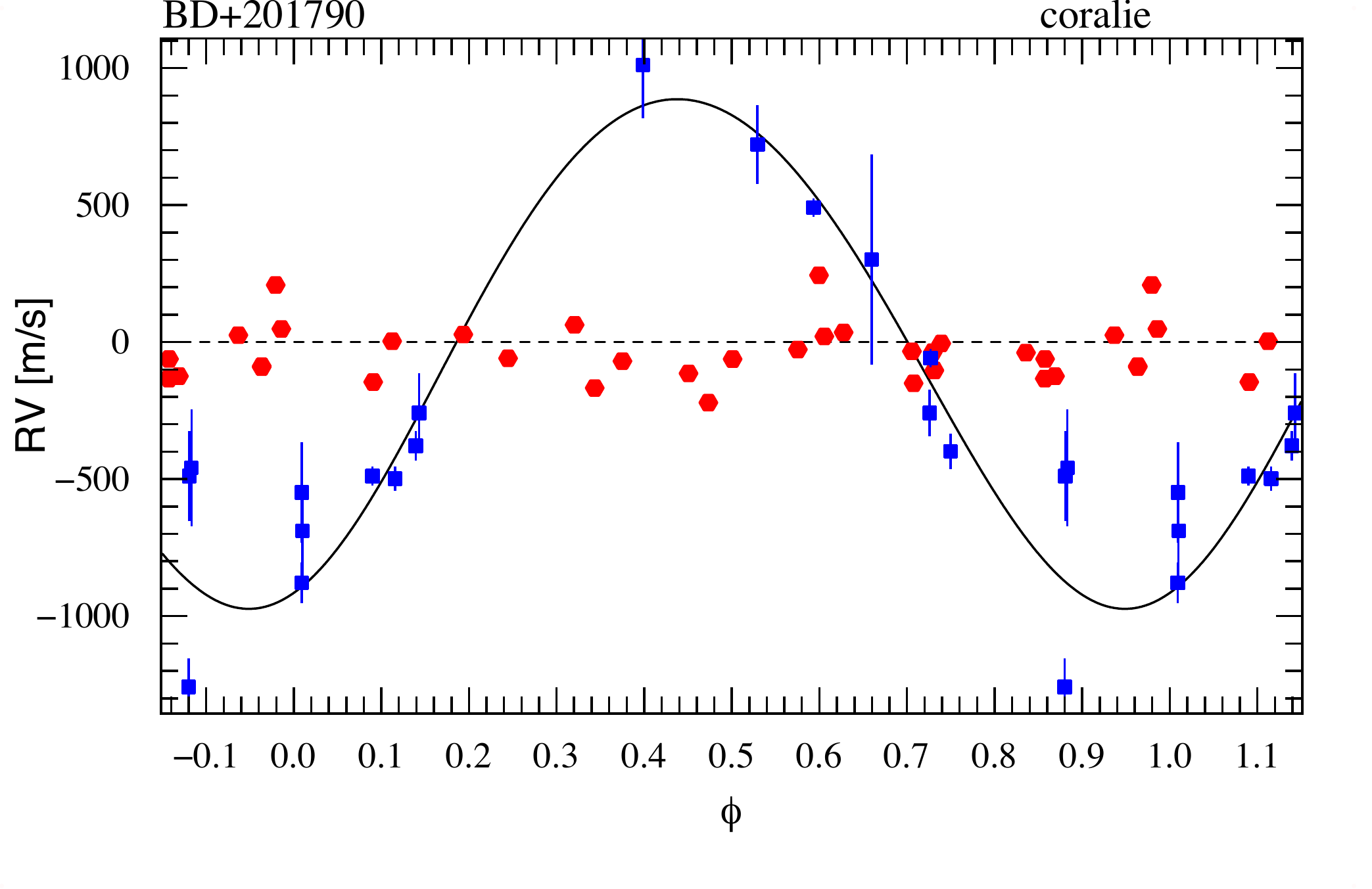}

\caption{CORALIE RV measurements for BD\,+20\,1270. The measured RVs are plotted as a function of time ({\it top pannel}) and phase-folded on the orbit announced by \cite{2009arXiv0912.2773H}, ({\it bottom panel}). The published RV measurements and error bars are presented as well ({\it square symbols}).}\label{RV_time}
\end{figure}

Our data points were obtained during two observing campaigns. The first set contains 20 measurements spanning 21 days, from the 21 December 2009 to the 10 January 2010; the second has 8 measurements obtained in a mission of 10 days, from the 4 to the 14 February 2010. The precision on the RVs is dominated by the photon noise contribution, with an average precision of 12.28\,m/s. The weighted r.m.s. of the data is of 101.7\,m/s and the peak-to-peak amplitude is of 464.9\,m/s. The data are presented in Fig. \ref{RV_time} (top panel).

The first point to note is the different amplitude of the RV variation on the two campaigns. On the first campaign the RVs are characterized by a weighted r.m.s. of 61.12\,m/s and a peak-to-peak amplitude is of 208.6\,m/s, while for the second these values are of 172.8\,m/s and 464.9\,m/s. The average uncertainties of the measurements are very similar, of 12.02 and 13.04\,m/s respectively, and cannot account for the discrepancy. In both cases the RV variations are smaller than the ones reported by \cite{2009arXiv0912.2773H} by a factor of 10 and 5, respectively, and cover a time span of three times the published orbital period for the first campaign, and once for the second. A phase-folded plot on the announced orbit depicts very well this discrepancy (Fig. \ref{RV_time}, bottom panel). We only considered the first of the two proposed orbits, but they only differ slightly and the same conclusions hold.

\begin{figure}
\includegraphics[width=9cm]{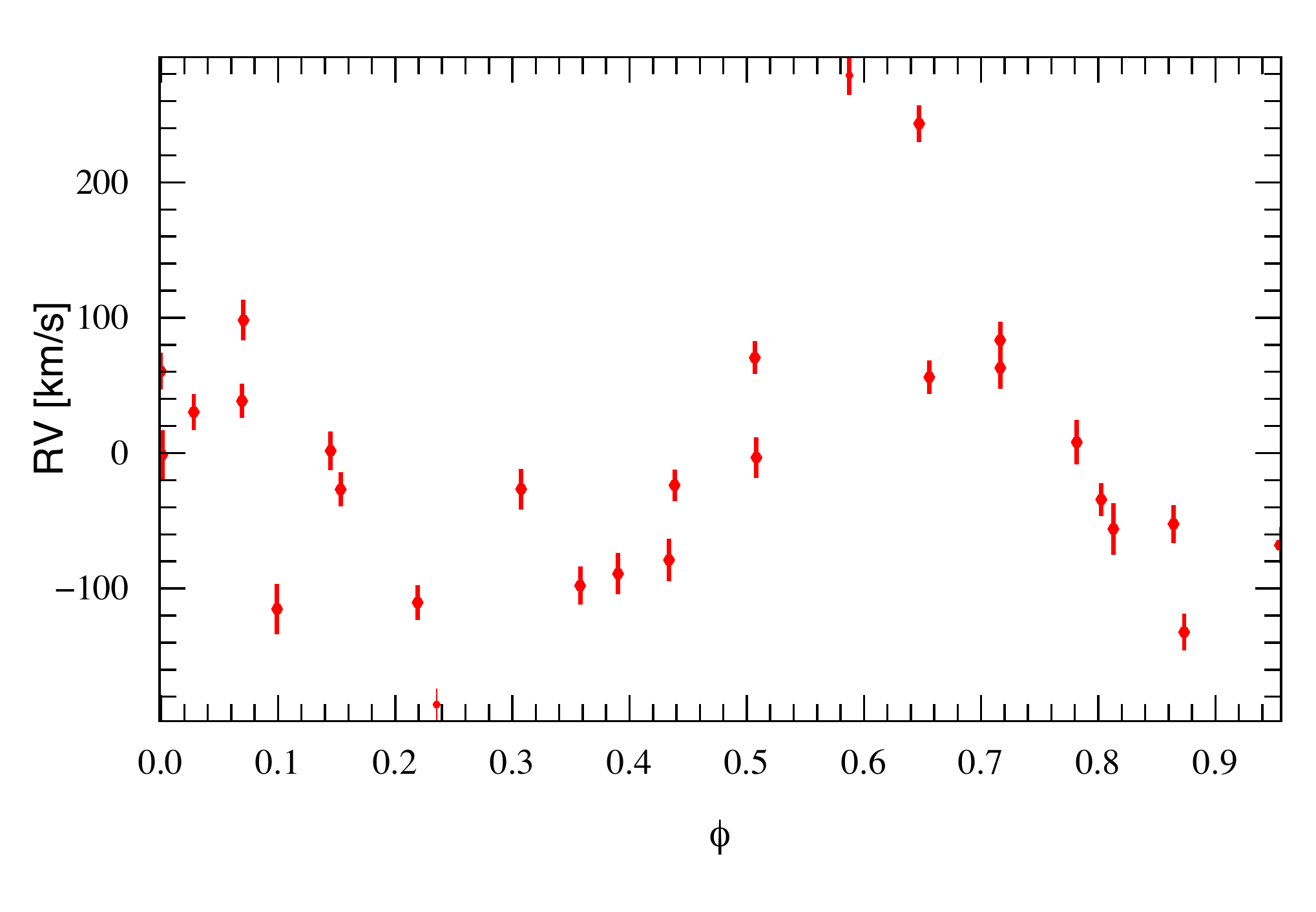}
\includegraphics[width=9cm]{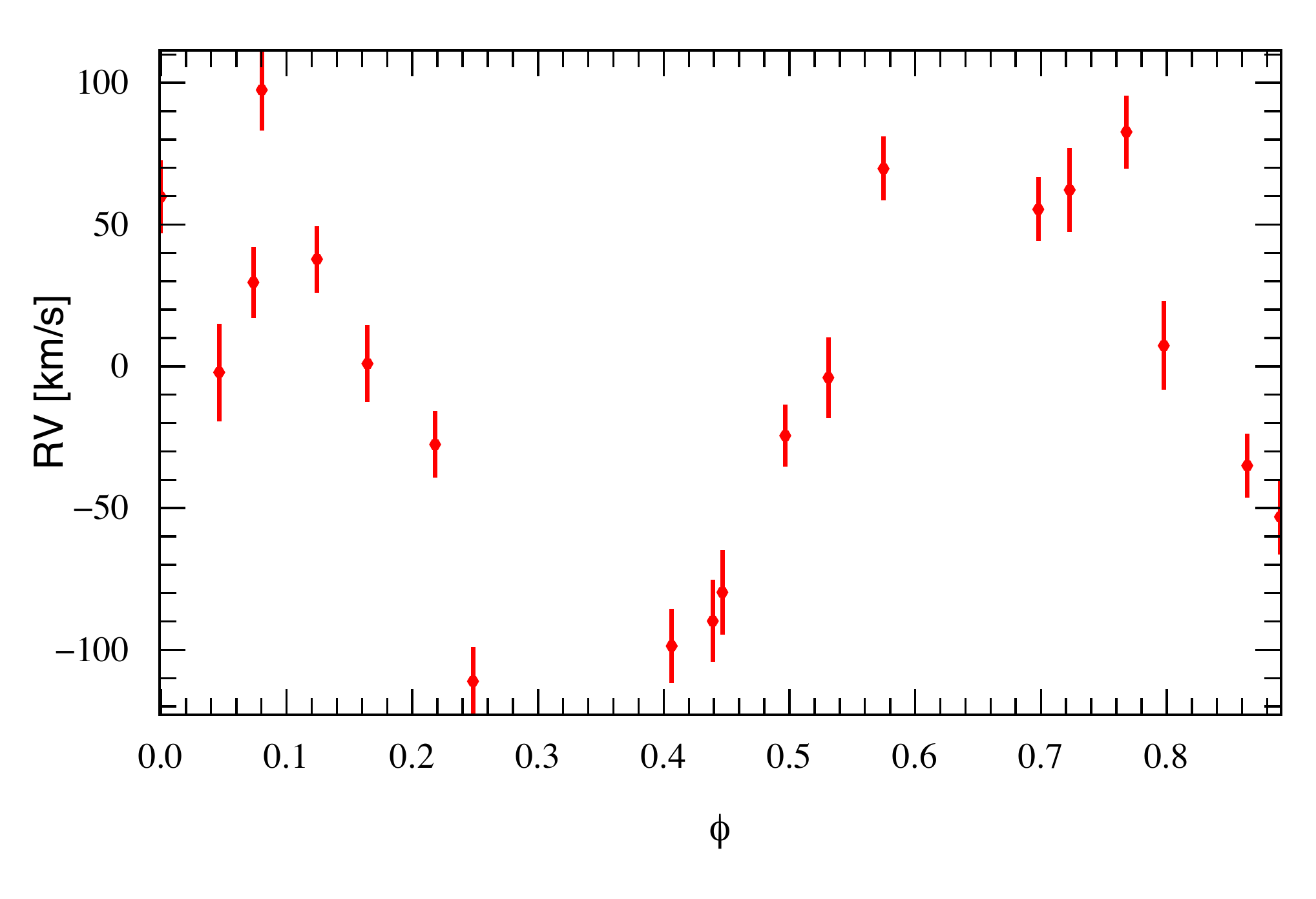}
\caption{CORALIE RV measurements for BD\,+20\,1270 folded on a 2.790 period ({\it top}), and the data from the first campaign only folded into a 2.765 period orbit ({\it bottom}). }\label{RV_folded}
\end{figure}

Still, the weighted r.m.s. is well in excess of the average measurement precision. In order to evaluate the presence of a periodic signal in the data we used two different approaches. The first was the ``string-length" method described in \cite{1983MNRAS.203..917D}. This method delivers the orbital period that minimizes the sum of the lengths of line segments in a (RV$_i$, $\phi_i$) diagram. It is suitable for randomly spaced observations in small data sets, such as ours, and is very efficient in detecting single planets in eccentric orbits. Using it we find a period of 2.790\,days (and a T$_0$ of JD\,=\,2455238.6). If only the data from the first campaign are used the period changes slightly, to 2.765 days. The phase-folded RV measurements for both data sets and periods are shown in Fig. \ref{RV_folded}. Note that while the two data sets show different RV amplitude variations, the periodicity of the signal is not significantly affected by the inclusion of the data from the second campaign. In particular, the data from the first campaign show a well-defined variation when phase-folded (bottom panel).

\begin{figure}
\includegraphics[width=9cm]{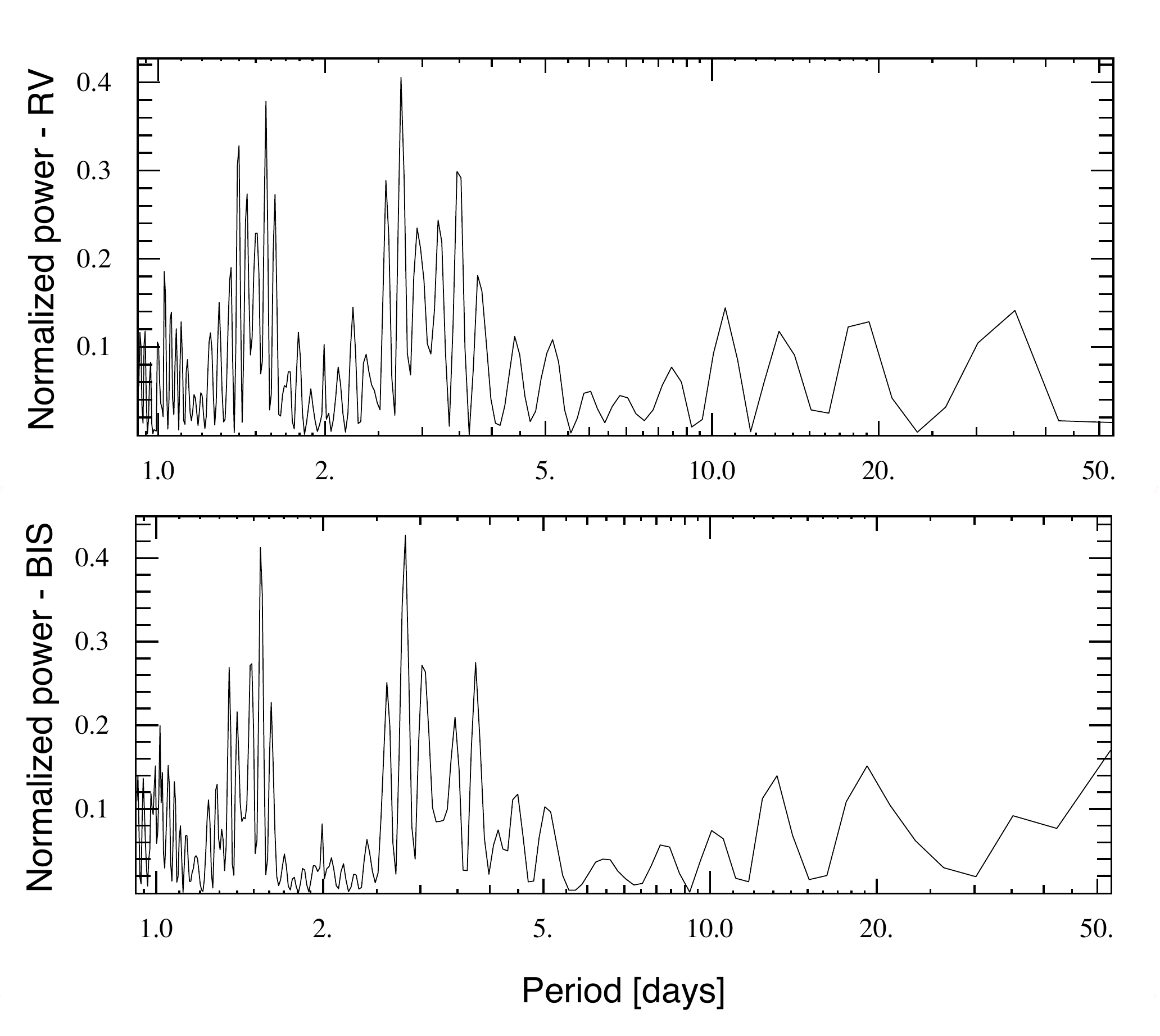}
\caption{Generalized Lomb-scargle periodogram for BD\,+20\,1790 RV ({\it top}), and BIS ({\it bottom}). }\label{GLS}
\end{figure}

\begin{figure}
\includegraphics[width=9cm]{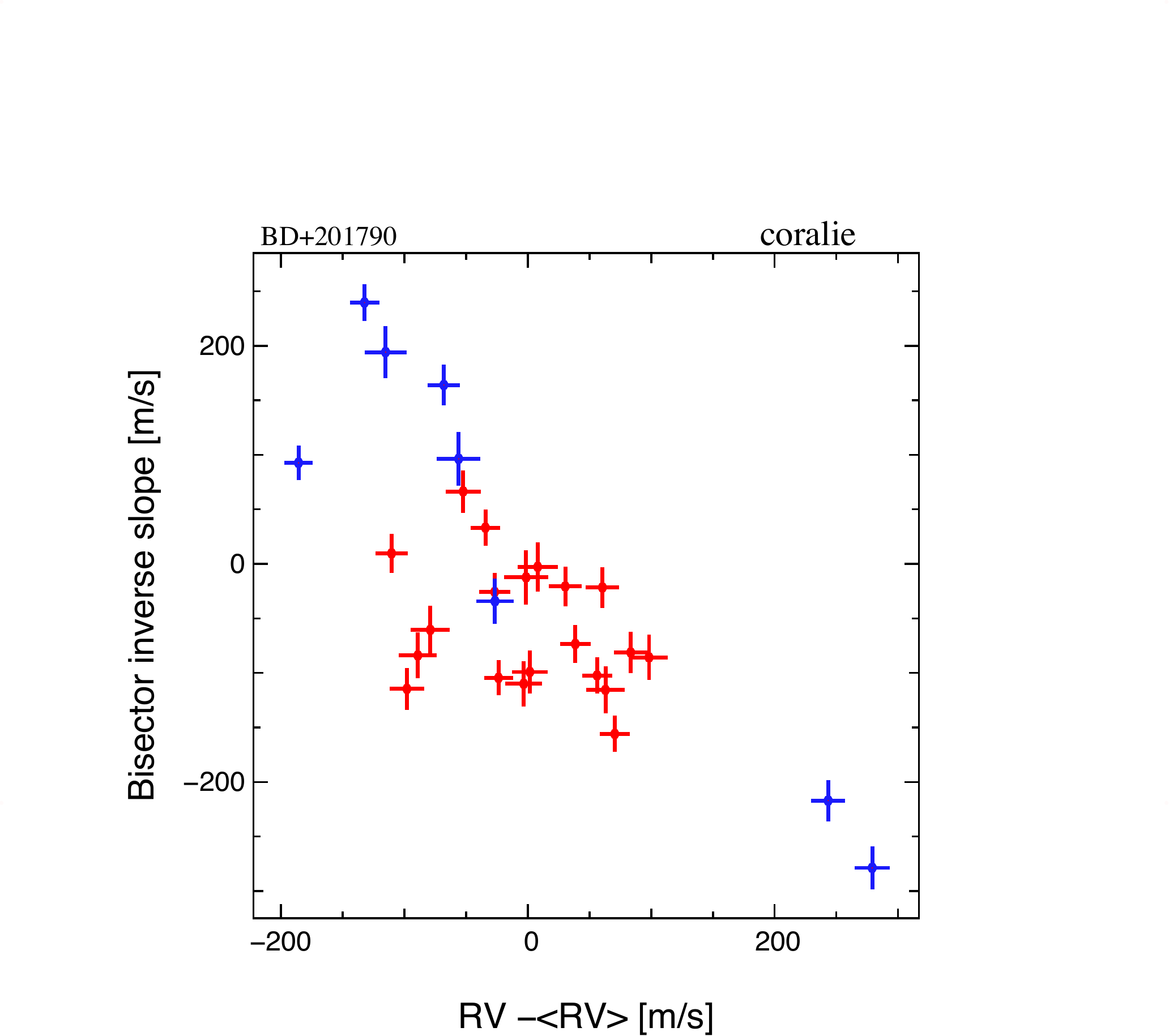}
\caption{ Correlation between RV and BIS. The red points correspond to the data from the first campaign and the blue points the data corresponding from the second ({\it electronic version only}). The BIS error bars are approximated by twice the photon error on the corresponding RV.}\label{BIS_RV}
\end{figure}

We also computed the generalized Lomb-Scargle periodogram \citep[as implemented by ][]{2009A&A...496..577Z}, which revealed the presence of a 1.55 and 2.8\,days signal in both the RVs and in the BIS (Fig. \ref{GLS}). The 1.55\,days signal is the alias of the 2.8\,days signal created by the 1\,day sampling. The BIS shows a significant variation, of the same order of the RV, and correlated with it. The BIS-RV plot is present in Fig.\ref{BIS_RV}; note that the data from the second campaign show extreme values of both RV and BIS. A linear least squares fit delivers a slope of -0.826. The Pearson's correlation factor is of -0.747, and Monte Carlo simulations show that the probability of obtaining this value or lower from two random non-correlated distributions with 28 points is $<$\,10$^{-4}$ and thus negligible.

\section{Discussion: Signal originated by a spot?}

In spite of our high-precision and temporal cadency of our RV measurements, we found no evidence for the planetary signal reported by \cite{2009arXiv0912.2773H}. Instead of an orbit with peak-to-peak amplitude of 1\,800\,m/s and period of 7.8\,days, we detected an RV signal with peak-to-peak variation of 460\,m/s and a periodicity of 2.8 days. Moreover, the amplitude of this signal is not constant, doubling from the first campaign to the second one. 

We tried to quantify the likelihood that our points were derived from the announced orbit. Since the error propagation on the T$_0$ makes the prediction of an accurate phase impossible, we proceeded as follows.  We considered a set of 28 data points randomly distributed over the entire phase, and calculated the RV for each point, and the r.m.s. of the set of measurements. Note that there is no instrumental error added to the curve, so this provides a lower limit to the r.m.s. The simulation was repeated 10\,000 times. All values were well in excess of the measured RV of 101\,m/s r.m.s., and we concluded then that the probability that the two data sets are compatible is lower than 10$^{-4}$.

Two arguments point towards a photospheric origin for the RV variations:

\begin{itemize}

\item the similarity between the RV periodicity found in the data and the announced photometric period;

\item the correlation between BIS and RV.

\end{itemize}

The hypothesis that the RV on BD+20\,1790 was rooted on stellar phenomena was discarded by \cite{2009arXiv0912.2773H}, who detected a RV signal with a periodicity different from the rotational period. Using rules of thumb from \cite{1997ApJ...485..319S} and \cite{2007A&A...473..983D}, the authors showed that the amplitude of the signal created by a spot was expected to be no larger than $\sim$600\,m/s. Even though this value is too small to explain their RV variation, it can easily account for what we detect. Moreover, the BIS on our measurements is significantly correlated with the RV, being particularly apparent that the data of the second campaign shows a much higher RV and BIS variation.




\section{Conclusions}

We present strong evidence that the RV variation on the star BD\,+20\,1790 is has its origin in the stellar atmosphere rather than being induced by the presence of an unseen companion. These conclusions were drawn from high-cadence, precise RV (at $\sim$10\,m/s level) that correctly sampled the previously proposed orbit. Instead of reproducing this orbit, we detected a lower RV variation with variable amplitude. The RV signal is correlated with BIS and shows a periodicity very similar to the reported photometric period.

This work shows the importance of correctly sampling the phase of a candidate orbit. Otherwise the conjugated effect of starspots and stellar jitter can be mistaken for a planetary signature.

\begin{acknowledgements}
     Support from the Funda\c{c}\~{a}o para Ci\^{e}ncia e a Tecnologia (Portugal) to P. F. in the form of a scholarship (reference SFRH/BD/21502/2005) is gratefully acknowledged.  NCS would like to acknowledge the support by the European Research Council/European Community under the FP7 through a Starting Grant, as well as from Funda\c{c}\~ao para a Ci\^encia e a Tecnologia (FCT), Portugal, through program Ci\^encia\,2007, and in the form of grants reference PTDC/CTE-AST/098528/2008 and PTDC/CTE-AST/098604/2008.
\end{acknowledgements}

\bibliographystyle{aa} 
\bibliography{Mybibliog} 

\begin{thebibliography}{14}
\expandafter\ifx\csname natexlab\endcsname\relax\def\natexlab#1{#1}\fi

\bibitem[{{Baranne} {et~al.}(1996){Baranne}, {Queloz}, {Mayor}, {Adrianzyk},
  {Knispel}, {Kohler}, {Lacroix}, {Meunier}, {Rimbaud}, \&
  {Vin}}]{1996A&AS..119..373B}
{Baranne}, A., {Queloz}, D., {Mayor}, M., {et~al.} 1996, \aaps, 119, 373

\bibitem[{{Desort} {et~al.}(2007){Desort}, {Lagrange}, {Galland}, {Udry}, \&
  {Mayor}}]{2007A&A...473..983D}
{Desort}, M., {Lagrange}, A.-M., {Galland}, F., {Udry}, S., \& {Mayor}, M.
  2007, \aap, 473, 983

\bibitem[{{Dworetsky}(1983)}]{1983MNRAS.203..917D}
{Dworetsky}, M.~M. 1983, \mnras, 203, 917

\bibitem[{{Hern{\'a}n-Obispo} {et~al.}(2009){Hern{\'a}n-Obispo},
  {G{\'a}lvez-Ortiz}, {Anglada-Escud{\'e}}, {Kane}, {Barnes}, {de Castro}, \&
  {Cornide}}]{2009arXiv0912.2773H}
{Hern{\'a}n-Obispo}, M., {G{\'a}lvez-Ortiz}, M.~C., {Anglada-Escud{\'e}}, G.,
  {et~al.} 2009, ArXiv e-prints

\bibitem[{{Hu{\'e}lamo} {et~al.}(2008){Hu{\'e}lamo}, {Figueira}, {Bonfils},
  {Santos}, {Pepe}, {Gillon}, {Azevedo}, {Barman}, {Fern{\'a}ndez}, {di Folco},
  {Guenther}, {Lovis}, {Melo}, {Queloz}, \& {Udry}}]{2008A&A...489L...9H}
{Hu{\'e}lamo}, N., {Figueira}, P., {Bonfils}, X., {et~al.} 2008, \aap, 489, L9

\bibitem[{{Mayor} \& {Queloz}(1995)}]{1995Natur.378..355M}
{Mayor}, M. \& {Queloz}, D. 1995, \nat, 378, 355

\bibitem[{{Pepe} {et~al.}(2002){Pepe}, {Mayor}, {Galland}, {Naef}, {Queloz},
  {Santos}, {Udry}, \& {Burnet}}]{2002A&A...388..632P}
{Pepe}, F., {Mayor}, M., {Galland}, F., {et~al.} 2002, \aap, 388, 632

\bibitem[{{Prato} {et~al.}(2008){Prato}, {Huerta}, {Johns-Krull}, {Mahmud},
  {Jaffe}, \& {Hartigan}}]{2008ApJ...687L.103P}
{Prato}, L., {Huerta}, M., {Johns-Krull}, C.~M., {et~al.} 2008, \apjl, 687,
  L103

\bibitem[{{Queloz} {et~al.}(2001){Queloz}, {Henry}, {Sivan}, {Baliunas},
  {Beuzit}, {Donahue}, {Mayor}, {Naef}, {Perrier}, \&
  {Udry}}]{2001A&A...379..279Q}
{Queloz}, D., {Henry}, G.~W., {Sivan}, J.~P., {et~al.} 2001, \aap, 379, 279

\bibitem[{{Saar} \& {Donahue}(1997)}]{1997ApJ...485..319S}
{Saar}, S.~H. \& {Donahue}, R.~A. 1997, \apj, 485, 319

\bibitem[{{Segransan} {et~al.}(2009){Segransan}, {Udry}, {Mayor}, {Naef},
  {Pepe}, {Queloz}, {Santos}, {Demory}, {Figueira}, {Gillon}, {Marmier},
  {Megevand}, {Sosnowska}, {Tamuz}, \& {Triaud}}]{2009arXiv0908.1479S}
{Segransan}, D., {Udry}, S., {Mayor}, M., {et~al.} 2009, ArXiv e-prints

\bibitem[{{Setiawan} {et~al.}(2007){Setiawan}, {Weise}, {Henning}, {Launhardt},
  {M{\"u}ller}, \& {Rodmann}}]{2007ApJ...660L.145S}
{Setiawan}, J., {Weise}, P., {Henning}, T., {et~al.} 2007, \apjl, 660, L145

\bibitem[{{Udry} \& {Santos}(2007)}]{2007ARA&A..45..397U}
{Udry}, S. \& {Santos}, N.~C. 2007, \araa, 45, 397

\bibitem[{{Zechmeister} \& {K{\"u}rster}(2009)}]{2009A&A...496..577Z}
{Zechmeister}, M. \& {K{\"u}rster}, M. 2009, \aap, 496, 577

\end{thebibliography}

\end{document}